\documentclass[a4paper,fleqn,usenatbib]{mnras}

\usepackage[T1]{fontenc}
\usepackage{ae,aecompl}

\usepackage{amsmath}
\usepackage{graphicx}
\usepackage{amssymb}

\renewcommand{\v}[1]{{\boldsymbol{#1}}} 

\newcommand{\uv}[1]{\hat{{\boldsymbol{#1}}}} 
\newcommand{\abs}[1]{\left| #1 \right|} 
\newcommand{\pd}[2]{\frac{\partial #1}{\partial #2}}
\newcommand{\curl}[1]{\v{\nabla} \times #1} 
\newcommand{\avg}[1]{\langle #1 \rangle} 

\title[Stellar Coronal Response to Flux Transport]{Stellar Coronal Response to Differential Rotation and Flux Emergence}

\author[G. P. S. Gibb, D. H. Mackay, M. M. Jardine, A. R. Yeates]{G. P. S. Gibb$^1$\thanks{E-mail: gpsg@st-andrews.ac.uk}, D. H. Mackay$^1$, M. M. Jardine$^2$, A. R. Yeates$^3$\\
$^{1}$School of Mathematics and Statistics, University of St Andrews, St Andrews, KY16 9SS\\
$^{2}$School of Physics and Astronomy, University of St Andrews, St Andrews, KY16 9SS\\
$^{3}$Department of Mathematical Sciences, Durham University, Durham, DH1 3LE}

\date{Accepted ?. Received ?; in original form ?}

\pubyear{2015}

\begin{document}
\label{firstpage}
\pagerange{\pageref{firstpage}--\pageref{lastpage}}
\maketitle

    \begin{abstract}
  We perform a numerical parameter study to determine what effect varying differential rotation and flux emergence has on a star's non-potential coronal magnetic field. In particular we consider the effects on the star's surface magnetic flux, open magnetic flux, mean azimuthal field strength, coronal free magnetic energy, coronal heating and flux rope eruptions. To do this, we apply a magnetic flux transport model to describe the photospheric evolution, and couple this to the non-potential coronal evolution using a magnetofrictional technique. A flux emergence model is applied to add new magnetic flux onto the photosphere and into the corona. The parameters of this flux emergence model are derived from the solar flux emergence profile, however the rate of emergence can be increased to represent higher flux emergence rates than the Sun's. Overall we find that flux emergence has a greater effect on the non-potential coronal properties compared to differential rotation, with all the aforementioned properties increasing with increasing flux emergence rate. Although differential rotation has a lesser effect on the overall coronal properties compared to flux emergence, varying differential rotation does alter the coronal structure. As the differential rotation rate increases, the corona becomes more open, and more non-potential.
    \end{abstract}


\section{Introduction}

Virtually all cool, partially-convective main sequence stars display evidence of hot, magnetically confined coronae. Evidence of hot coronae mainly consists of X-ray emission \citep{Vaiana1981,Schmitt1997,Feigelson1999}. In addition to this, evidence of magnetic fields on such stars include: the presence of dark spots (interpreted to be analogous to sunspots) found through tomographical techniques \citep{Vogt1983,Strassmeier1996,Berdyugina2005}, rotational modulation \citep{Radick1982,Radick1983} and even exoplanetary transits \citep{Rabus2009,Pont2007,Sanchis-Ojeda2011}. Additionally, through the Zeeman effect, direct measurements of photospheric magnetic fields have been made \citep{Robinson1980}, and using Zeeman Doppler Imaging (ZDI), maps of the photospheric magnetic fields of stars can be constructed \citep{Semel1989,Brown1991,Donati1997,DonatiCameron1997}. Periodic variability of the chromospheric emission of stars over timescales of years \citep{Wilson1978} and long term variability in X-ray emission \citep{Favata2008,Ayres2009} are interpreted to be due to magnetic cycles analogous to the Sun's eleven year cycle. These observations add further evidence to the argument that the coronae are magnetically driven.

A number of studies have been conducted that have determined the response of the coronae to various stellar parameters. The X-ray luminosity is found to increase with increasing stellar rotation period \citep{Pallavicini1981,Walter1981}, as do chromospheric emission proxies \citep{Middelkoop1981,Mekkaden1985}. Similarly, the mean surface field strength, $\avg{B}$, is found to increase with decreasing rotation period \citep{Vidotto2014}. In addition to the activity-rotation relation, the activity of stars is also dependent upon stellar mass. For a set of stars with similar rotation periods, the activity of stars increases with decreasing mass (e.g. \citet{Donati22009}, \citet{Marsden2014}).

On the Sun, the corona may largely be considered to be in equilibrium, save for impulsive events such as flares and coronal mass ejections (CMEs) \citep{Priest1982}. The large-scale coronal evolution is thus predominantly driven through the evolution of the photospheric magnetic field, in which the coronal field is anchored. The photospheric magnetic field evolves in response to large-scale surface flows (such as differential rotation and meridional flow) and small-scale convective flows (granulation and supergranulation). Differential rotation is the tendency for a star's equator to rotate faster than its poles. This has the effect of stretching magnetic features out in the east-west direction, and builds up shear in the corona \citep{vanballegooijen2000,mackay2006a}. Meridional flow is a polewards flow which acts to migrate magnetic features towards higher latitudes \citep{Babcock1961,DeVore1985}. Granulation and supergranulation act to jostle small-scale magnetic elements about, and break up large magnetic field distributions. The large-scale effect of granulation on the magnetic field can be treated as a diffusion process \citep{Leighton1964}. For a review on the understanding of surface magnetic flux transport processes on the Sun, please see \citet{Sheeley2005} or \citet{MackayYeates2012}.

In addition to surface flows, the coronal field is also altered through the addition of new magnetic flux, which emerges through the photosphere from the solar interior. Flux emergence is thought to be due to magnetically buoyant flux tubes rising up through the convective zone (e.g. \citet{Fan2001}), and typically results in bipolar active regions at the photosphere. Flux emergence rejuvenates the corona, by adding new flux into the corona, and altering its structure. On the Sun, active regions preferentially emerge between $\pm 35^\circ$ latitude \citep{Priest1982}, are predominantly tilted such that the leading polarity (in the direction of rotation) is closer to the equator \citep{Hale1919}, and the sign of the leading polarity in one hemisphere is opposite to the sign of the leading polarity in the other hemisphere \citep{Hale1919}.

On stars, differential rotation may be estimated by considering the rotational modulation of chromospheric emission lines over a stellar activity cycle \citep{Donahue1996}. More recently, Doppler Imaging and ZDI have been able to better measure differential rotation \citep{Donati1997,Petit2002,Donati2000,Marsden2006,Waite2011}. Such studies have found differential rotation rates of up to eight times the solar value (e.g. \citet{Marsden2011}), and it is in general found that the differential rotation rates of stars increase with increasing stellar effective temperature \citep{Barnes2005,Cameron2007,Kuker2011}. At present, flux emergence rates cannot be measured for stars other than the Sun.

In our previous paper \citep{Gibb2014-2} we investigated the effects of differential rotation and supergranulation (parameterised through a surface diffusion) on coronal timescales. To do this we modelled the coronal evolution of a single isolated bipolar active region on a portion of a star. The differential rotation facilitated the shearing of the coronal field, and the surface diffusion allowed flux cancellation, resulting in the transformation of sheared field into a flux rope. The differential rotation eventually initiated the eruption of the flux rope. We found that the formation times and lifetimes of the flux ropes decreased with increasing differential rotation rate. We interpreted this to mean that the coronal timescales on stars decrease with increasing differential rotation, and postulated that high differential rotation stars may have more frequent eruptions and flares compared to low differential rotation stars. We also discussed that the timescales deduced from the work may not be reflected in more complicated circumstances, as active regions are unlikely to be isolated, but rather are likely to be in proximity to other active regions. The proximity to other active regions may affect the timescales through mutual interactions between neighbouring active regions' coronal fields.

In this paper we carry out a numerical parameter study to investigate the effects of varying the differential rotation and flux emergence rates on a number of global coronal and photospheric properties. In Section \ref{model} we describe the photospheric and non-potential coronal evolution model we employ. Section \ref{bipole_emergence_model} describes the construction of our flux emergence model. Our results are presented in Section \ref{results} and finally we summarise our results and discuss them in Section \ref{discussion}.

\section{Numerical Model} \label{model}
In this study we simulate the evolution of stars' non-potential coronal magnetic fields in response to the evolution of their surface magnetic fields. The surface evolution is driven by a surface flux transport model, which includes the effects of differential rotation, meridional flow and surface diffusion (see Section \ref{flux_transport_model}). The non-potential coronal evolution is achieved through a magnetofrictional technique (Section \ref{magnetofriction}) which produces a series of non-linear force free fields. In addition to these, we also include flux emergence into the model to account for the addition of new magnetic flux onto the photosphere and into the corona over the duration of the simulations (Section \ref{flux_emergence_model}). The photospheric and coronal evolution are treated in the same manner as \citet{Gibb2014-2}, with the exception of the flux emergence model, which uses the method of \citet{Yeates2008}.

We employ a numerical grid using the variables $(x,y,z)$, related to $(r,\theta,\phi)$, the radius, co-latitude {(related to the latitude, $\lambda$, by $\theta^\circ = 90^\circ - \lambda$)}  and longitude by:
\begin{align}
x &= \frac{\phi}{\Delta}, \\
y  &= -\frac{\ln \left(\tan \frac{\theta}{2}\right)}{\Delta}, \\
z  &= \frac{\ln \left(\frac{r}{R_*}\right)}{\Delta},
\end{align}
where $\Delta$ is the grid spacing. This choice of variables ensures that the horizontal cell size is $h_\phi = h_\theta = r \Delta \sin{\theta}$ and the vertical cell size is $h_z = r\Delta$. A variable resolution grid is employed in order to prevent the cell sizes from becoming too small towards the poles. At the equator the grid spacing is $0.9375^\circ$ whilst at the poles the grid spacing is $30^\circ$.  For more details on the variable-resolution grid employed, please refer to \citet{Yeates2014}.
A staggered grid is applied in order to achieve second order accuracy for the computation of derivatives.
{The corona is simulated at all longitudes, between $\pm 89.5^\circ$ latitude and between the stellar surface and $2.5R_*$.}
We apply closed boundary conditions on the latitudinal boundaries{, and periodic boundary conditions on the longitudinal boundaries}. At the upper ($r=2.5R_*$) boundary we apply an open boundary condition, and the lower ($r=R_*$) boundary is specified by the radial photospheric magnetic field as deduced from the flux transport model described below.

\subsection{Surface Flux Transport Model} \label{flux_transport_model}
In order to model the evolution of the non-potential coronal magnetic field with the magnetofrictional method, we require a description of the evolution of the photospheric magnetic field. The photospheric evolution is determined using the flux transport model described in \citet{Gibb2014-2}. This model assumes the radial photospheric magnetic field, ${B}_r$, is influenced through the effects of differential rotation, meridional flows and surface diffusion. We express the radial magnetic field at the photosphere by the vector magnetic potentials $A_\theta$ and $A_\phi$ through
\begin{equation}
B_r = \frac{1}{r\sin{\theta}}\left[\pd{}{\theta}(\sin{\theta}A_\phi) -\pd{A_\theta}{\phi}\right].
\end{equation}
The radial photospheric field is evolved by solving the two dimensional flux transport equation:
\begin{align}
\pd{A_\theta}{t} &= u_\phi B_r - \frac{D}{r\sin{\theta}}\pd{B_r}{\phi} ,\\
\pd{A_\phi}{t} &= -u_\theta B_r + \frac{D}{r}\pd{B_r}{\theta},
\end{align}
where $u_\phi$ is the azimuthal velocity, $u_\theta$ is the meridional flow velocity and $D$ is the photospheric diffusion constant, taken in this study to be $450$ km$^2$s$^{-1}$. {It should be noted that although in this study we employ the magnetic vector potential, $\v{A}$, it is also possible to use a `superpotential' for the poloidal field, akin to the methods used in dynamo models (e.g. \citet{Moffatt1978,Krause1980,Bigazzi2004}).}

We take the azimuthal velocity to be
\begin{equation}
u_\phi = \Omega(\theta)r\sin{\theta},
\end{equation}
where
\begin{equation} \label{domega scaling}
\Omega(\theta) = K\left(\Omega_0 - d\Omega_\odot \cos^2{\theta}\right) \text{ deg day}^{-1}.
\end{equation}
The term $\Omega (\theta)$ is the angular velocity of rotation relative to the rotation at $30^\circ$ latitude ($60^\circ$ co-latitude). We choose $\Omega_0 = 0.9215\text{ deg day}^{-1}$ and $d\Omega_\odot = 3.65\text{ deg day}^{-1}$ to represent the solar differential rotation profile. The constant, $K$ scales the differential rotation to that of other stars. For example, if $K=2$ the differential rotation rate will be twice that of the solar rate.

The meridional velocity is prescribed by
\begin{equation}
u_\theta = C \frac{16}{110}\sin(2\lambda)\exp(\pi-2\abs{\lambda}),
\end{equation}
where $\lambda = \pi/2 - \theta$ is the latitude and $C=11 \text{ ms}^{-1}$ is the solar peak meridional flow speed. It is important to note that although we have prescribed the solar meridional flow speed, a study by \citet{Mackay2004} suggested that some stars may have meridional flow speeds on the order of ten times the solar value. In this work the effect of the meridional flow speed on the coronal response is not investigated, however this will be considered in future work.

\subsection{Coronal Evolution Model} \label{magnetofriction}
We evolve the coronal magnetic field using the ideal induction equation,
\begin{equation}
\pd{\v{A}}{t} = \v{v} \times \v{B},
\end{equation}
where $\v{B} = \curl{\v{A}}$ and the velocity contains contributions from the magnetofrictional velocity ($\v{v}_{\text{MF}}$) and an outflow velocity ($\v{v}_\text{out}$), both of which are described below. We employ the magnetic vector potential, $\v{A}$, as the primary variable in this study as its use in conjunction with a staggered grid ensures the solenoidal constraint is met. {It should be noted that in this study we do not explicitly include diffusion in our simulations. Whilst this is the case, numerical diffusion allows processes such as reconnection to occur.}

In the magnetofrictional approach \citep{Yang1986} the velocity in the induction equation is set to be
\begin{equation} \label{mfvel}
\v{v}_\text{MF} =  \frac{1}{\nu} \frac{\v{j}\times\v{B}}{B^2},
\end{equation}
such that it is parallel to the Lorentz force. The changing photospheric magnetic field -- as specified by the flux transport model -- induces a Lorentz force above the photosphere. The magnetofrictional velocity, which is aligned in the direction of this Lorentz force, acts to advect the coronal field towards a new non-linear force-free equilibrium.
The changing photospheric field thus drives the evolution of the coronal field through a series of force-free equilibria based upon the evolution of the photospheric field.

We apply a radial outflow velocity to ensure that the coronal magnetic field at the upper boundary is radial, and also to allow any flux ropes that have lifted off from the photosphere to leave the computational box. The outflow velocity is taken to be
\begin{equation}
\v{v}_\text{out} = v_0 \exp\left(\frac{r-2.5R_{*}}{r_w}\right) \uv{r},
\end{equation}
where $v_0 = 100 \text{ km s}^{-1}$ and $r_w$ is the width over which the radial velocity falls off at the outer boundary. Once the field lines become radial ($\v{B} = B_r\uv{r}$) near the outer boundary the outflow velocity has no effect on the evolution of the magnetic field as $\v{v}_\text{out} \times \v{B} = 0$.

\subsection{Flux Emergence} \label{flux_emergence_model}
We treat the mechanism of flux emergence by inserting idealised bipoles which represent active regions at the times and locations as specified by our bipole emergence model (see Section \ref{bipole_emergence_model}). The bipole emergence is achieved in the same way as that used by \citet{Yeates2008}, and will now be summarised.
The inserted bipole's vector potentials take on the following form:
\begin{align}
A_x =& \beta B_0e^{0.5}z \exp(-2\xi) \label{proj3_bipole1} \\
A_y =& B_0 e^{0.5}\rho \exp(-\xi) \\
A_z =& -\beta B_0 e^{0.5}x' \exp(-2\xi) \label{proj3_bipole3}
\end{align}
where
\begin{equation}
\xi = \frac{(x'^2+z^2)/2 + y'^2}{\rho^2},
\end{equation}
\begin{align}
x' &= x\cos ({-\gamma}) + y\sin({-\gamma}) \\
y' &= y\cos ({-\gamma}) - x\sin({-\gamma}),
\end{align}
\begin{equation}
B_0 = \frac{\Phi}{\sqrt{\pi e} \rho^2}.
\end{equation}
The term $\rho$ is the bipole's half separation, $\Phi$ is the bipole's flux and $\gamma$ is the bipole's tilt angle. The parameter $\beta$ describes the amount of twist the emerged bipole's field contains (see \citet{Yeates2009} for further details). Please note that the sign of $B_0$ (and thus the flux) may be positive or negative, and specifies the sign of the trailing polarity in the bipole. Also note that the above equations produce a bipole at the equator. This bipole is then rotated to the correct position ($x_0,y_0$) on the sphere.

Before the bipole is inserted any pre-existing coronal (and photospheric) magnetic field within the volume to be occupied by the bipole must be swept away. This is to ensure that the addition of the new field does not lead to any disconnected flux in the corona. It also mimics the distortion of pre-existing coronal field by the newly emerging flux, as has been observed in simulations of flux emergence \citep{Yokoyama1996,Krall1998}. In order to carry out the bipole insertion, the following steps are carried out:
\begin{enumerate}
\item The simulation's time is `frozen' by switching off the surface flows, surface diffusivity and the magntofrictional and outflow velocities.
\item An outward velocity is applied in a dome centred upon the new bipole's position to sweep the pre-existing coronal/photospheric field away.
\item The bipole is inserted into the corona by adding the bipole's vector potential (Equations \ref{proj3_bipole1} to \ref{proj3_bipole3}) to the pre-existing coronal vector potentials.
\item The magnetofrictional and radial outflow velocities are switched on, and the new bipole is allowed to reach an equilibrium with its surroundings for 50 timesteps.
\item Time is restarted by turning on the photospheric flows and diffusion.
\end{enumerate}
For further details of the flux emergence method applied, please see \citet{Yeates2008}.

\section{Bipole Emergence Model} \label{bipole_emergence_model}

At the present time stellar flux emergence profiles are not known, however, the nature of stellar emergence profiles has been speculated \citep{Mackay2004} and methods have been developed that may in the future be able to constrain them (e.g. \citet{Llama2012}, and see \citep{Berdyugina2005} for a review). As there is very little information available on the flux emergence profiles on stars, we base our stellar flux emergence profile on the well known solar flux emergence profile.
On the Sun, the flux emergence profile is time dependent. The butterfly diagram is an illustration of this time dependence. Although activity and magnetic cycles have been observed (e.g. \citet{Wilson1978,Donati2008,Fares2009}), in this work we are interested in the steady state corona, and so we wish to construct a flux emergence profile whose parameters do not vary significantly in time. In other words, we wish to simulate a time period sufficiently shorter than the stellar cycle's period so that the emergence profile's parameters are nearly independent with time. Doing so allows us to more clearly identify how varying the differential rotation and flux emergence rates effect the coronal field.

To construct our flux emergence profile, we consider the emergence of flux on the Sun between January 2000 and January 2001. During this time the distribution of the properties of emerged flux is approximately time-independent, and this time range approximately coincides with solar maximum. We use the properties of the flux that was determined to have emerged during this time period by \citet{Yeates2014}. The properties are derived from synoptic magnetograms of the Sun, and determine the locations of newly emerged flux. The newly emerged flux regions were approximated to be bipoles, whose fluxes, latitudes, longitudes, tilt angles, emergence times and half separations were chosen to best represent the observed emerged flux. For further details on how these values were obtained, see \citet{YeatesMackayvanBallegooijen2008}. In the year between January 2000 and January 2001, \citet{Yeates2014} determined that 227 bipoles had emerged. In this section, we describe a bipole emergence model that is able to reproduce the distribution of the bipoles as found by \citet{Yeates2014}. We chose to develop a model rather than just reuse the values obtained by \citet{Yeates2014} so that we can vary the parameters of the model to change the flux emergence rate.

\subsection{Observed Solar Profile}
We now describe the set of empirical relations obtained from the properties of emerged bipoles on the Sun between January 2000 and January 2001. Over this time period, 227 bipoles emerged, and the time interval between individual emergences was found to follow an exponential distribution,
\begin{equation}
	P(t)=\frac{1}{\tau} \exp\left(\frac{-t}{\tau}\right)
\end{equation}
with a mean interval, $\tau$, of 1.61 days. This corresponds to an emergence rate of 0.62 bipoles per day. Similarly, the bipole fluxes were found to follow an exponential distribution with a mean flux of $8.40\times10^{21}$ Mx.

Next, we determined the empirical scalings for the tilt angle with latitude. We found that
\begin{equation}
	\gamma^\circ \propto 0.32 \lambda^\circ \pm 15^\circ,
\end{equation}
where the quantity after the `$\pm$' denotes the standard deviation from the relation. Similarly, the relationship between the tilt angle and flux was found to be:
\begin{equation}
	\gamma^\circ \propto -0.32\Phi (\times 10^{21}\text{ Mx}) \pm 15^\circ,
\end{equation}

\begin{figure}
\centering\includegraphics[scale=0.5]{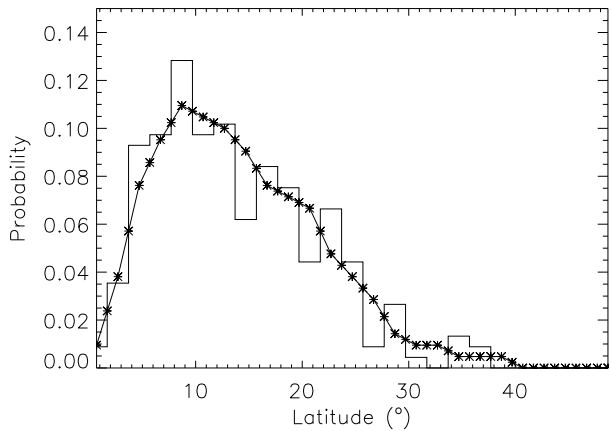}
\caption{Normalised histogram of the observed distribution of the unsigned latitudes of emergence. Also shown is the smoothed normalised histogram which has been resampled to a higher resolution (crosses and solid line) for use in constructing a synthetic distribution of latitudes based on the observed solar distribution. \label{emergence_lats}}
\end{figure}

We find the relationship between the separation and the flux to be approximately described by the piecewise function below:
\begin{equation} \label{proj3_sepeqn}
\rho = \left\{
\begin{array}{lr}
1.24+0.272\Phi \pm 0.5 & 0 < \Phi \le 12\times 10^{21} \text{ Mx} \\
4.51 \pm 1 & \Phi > 12\times 10^{21} \text{ Mx}
\end{array}
\right.
\end{equation}
The separations and tilt angles and the separations and latitudes showed no correlation. In the date range, it was found that 97\% of the bipoles obeyed Hale's law, whilst 3\% did not. We take the relationship between the beta and latitude to be
\begin{equation}
	\beta \propto -0.016 \lambda^\circ \pm 0.4
\end{equation}
to be consistent with the prescription of \citet{Yeates2010}. Lastly, the distribution of latitudes of emergence is plotted in Figure \ref{emergence_lats}.

\subsection{Model}

In the previous section we determined a series of relationships between the various bipole properties. With this information, we can now construct a bipole emergence model to represent the observed solar flux emergence profile between the years 2000 and 2001. This model can be used to consider stars with different flux emergence rates to the that of the Sun, by altering the mean time between emergences in the model. Below we outline the procedure used to obtain the bipole properties from the bipole emergence model. The procedure uses a random number generator where we use the shorthand $U[a,b]$ for a number drawn from the uniform distribution between $a$ and $b$, $N[\mu,\sigma]$ for a number drawn from a normal distribution with mean, $\mu$, and standard deviation, $\sigma$. Finally, we use $E[\tau]$ to denote a number drawn from an exponential distribution with expected value, $\tau$. Starting at $t=0$ days, the following procedure is carried out:

\begin{enumerate}

\item Determine the time for the next emergence, $\delta t$ from $E[1.61/ D \text{ days}]$, where $D$ is a scaling parameter to scale the flux emergence rate to be lower or higher than the Sun's. Set the time, $t$, to $t+\delta t$.

\item Fix the longitude of emergence from $U[0^\circ,360^\circ]$.

\item Select the latitude of emergence from the distribution shown in Figure \ref{emergence_lats}.

\item With a 50\% probability, choose the sign of the latitude to be negative to allow an approximately equal number of emergences in the northern and southern hemispheres of the star.

\item Determine the bipole flux from $E[8.40\times10^{20} \text{ Mx}]$.

\item With a probability of 97\%, set the flux to have the opposite sign as the latitude. Otherwise set the flux to have the same sign. This ensures that 97\% of the bipoles obey Hale's law, and the remainder do not.

\item Set the tilt angle to be $0.32 \lambda + N[0,14.93^\circ]$. The purpose of the normally distributed term is to add scatter to the tilt angles.

\item Set $\beta$ to be $-0.016\lambda + N[0,0.4]$. Once again, the purpose of the normally distributed term is to add scatter to $\beta$.

\item If the flux is less than $12\times10^{20}$ Mx then set the separation to be $1.24+0.272\Phi + U[-0.5,0.5]$, else set the separation to be $4.51+N[0,1]$. As above, the uniform and normally distributed random numbers are included to add scatter to the separations.

\item Repeat the procedure until $t \ge 365$ days.

\end{enumerate}

Although in the above model we did not specify the relations between $\rho$ and $\lambda$, $\rho$ and $\gamma$ and $\gamma$ and $\Phi$, the relations (or lack thereof) between these parameters as derived from our model match those of the Sun's in the sample we used to construct the model.
This demonstrates that our simple empirical bipole emergence model is able to reproduce a solar-like emergence profile with a minimal number of explicitly specified parameters.

\section{Results} \label{results}
In this section we describe the results from simulations run where we have varied both the flux emergence rate (i.e. the number of bipoles emerging per day) and the differential rotation rate (i.e. the equator-pole lap time). We investigate various combinations of $D=1$, 2, 3, 4 and 5 times the solar flux emergence rate of 0.62 bipoles per day, and $K=1$, 2, 3, 4 and 5 times the solar differential rotation rate of $3.65^\circ$ per day. The simulations consider the star's surface and coronal evolution over a period of one year. In particular we investigate the effects that the flux emergence and differential rotation have on `global' quantities, such as the open and surface flux, magnetic energy and electric currents. We also investigate the numbers of flux ropes formed in the simulations, and present $j^2$ emission proxy images of the stellar coronae, comparing them to field line plots.

\subsection{Simulation Set-Up}
\begin{figure}
\centering\includegraphics[scale=0.35]{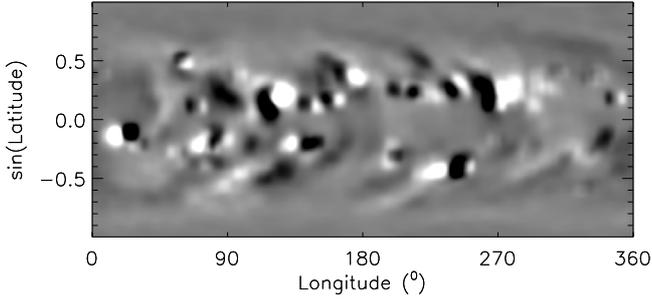}
\caption{The synoptic magnetogram used to construct the potential initial condition used in the simulations. The image is saturated at $\pm$50 G.\label{proj3_ic}}
\end{figure}

Rather than emerging the bipoles into an initially empty corona, we use a smoothed synoptic magnetogram of the solar photospheric field to construct a potential field initial condition for the coronal field. The synoptic magnetogram is from Carrington Rotation 1970, corresponding to the date range 24th Nov 2000 to 21st Dec 2000. This date range is from within the range of dates used to construct the bipole emergence model. The magnetogram's photospheric field distribution therefore has a spatial distribution of field consistent with our empirical flux emergence profile, albeit with the solar flux emergence rate. In Figure \ref{proj3_ic} the synoptic magnetogram from Carrington rotation 1970, used to construct the initial potential coronal field, is displayed.

\subsection{Simulation Behaviour}
\begin{figure}
\centering\includegraphics[scale=0.5]{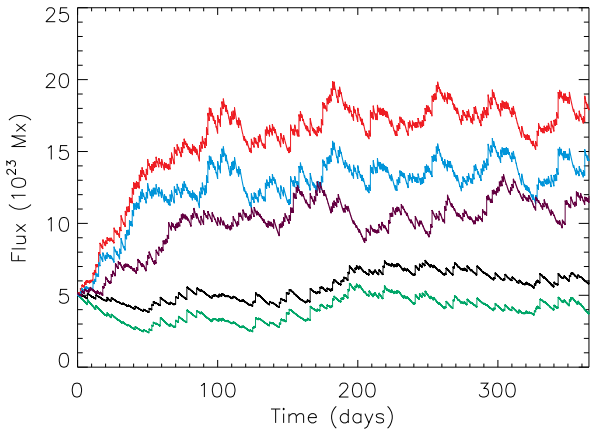}
\caption{The evolution of the surface flux for the simulations with $(D,K) = (1,1)$--black, $(5,1)$--red, $(1,5)$--green, $(5,5)$--blue and $(3,3)$--purple. Note that after 100-200 days the flux in each simulation reaches a steady state. \label{simulation_behaviour}}
\end{figure}
We will now briefly describe the general behaviour of each simulation. Over the course of any given simulation, the global quantities (e.g. flux, magnetic energy etc...) evolve in response to the surface flux transport and the addition of new flux. For the first 100--200 days the quantities either increase or decrease until they reach a near steady state. Once this steady state has been achieved, the quantities then vary by a certain amount around a mean value. To demonstrate this, Figure \ref{simulation_behaviour} displays the evolution of the surface flux with time for five of the simulations. In the sections below, when we refer to a quantity we refer to the mean, steady state value determined after $t=200$ days.

When a steady state has been achieved, the rate of input of that quantity into the simulation is approximately equal to the rate of dissipation of that quantity. For example, if the quantity in question was the magnetic energy, then the Poynting flux (input of energy) would be approximately equal to the losses due to magnetofrictional dissipation in the volume and energy leaving through the top boundary. As such, the initial increase/decrease in a given quantity is due to an imbalance in the injection/dissipation rates of the quantity.

\subsection{Magnetic Flux}
\begin{figure}
\centering\includegraphics[scale=0.5]{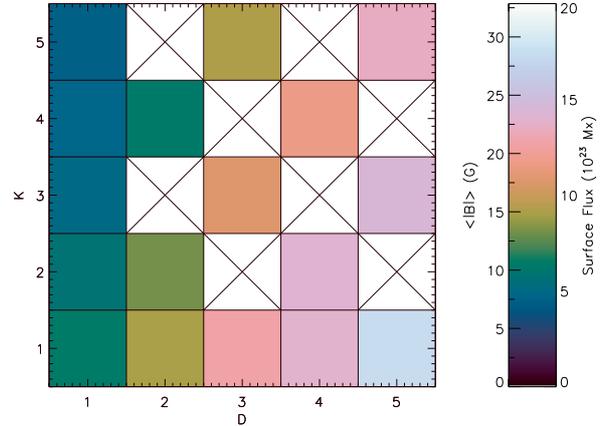}
\caption{The steady state surface fluxes for each simulation. Also displayed are the mean surface magnetic field strength. White squares {with crosses} contain no data (e.g. no simulation was run for that combination of $(D,K)$).\label{flux_fig}}
\end{figure}
To begin with, we will consider the magnetic flux, both the surface flux ($\Phi_s$), and the open flux ($\Phi_o$). Figure \ref{flux_fig} displays the steady state surface flux for each simulation run. Also displayed is the mean surface magnetic field strength, $\avg{|B|}$, calculated from
\begin{equation}
\avg{|B|} = \frac{\Phi_s}{\oint dS},
\end{equation}
where $dS$ is an infinitesimal element of the stellar surface. At a glance it can be seen that the surface flux and mean field increase with increasing flux emergence rate, and decrease with increasing differential rotation rate. The increase in flux with increasing flux emergence rate is due to the increased flux emergence rate adding more flux -- on average -- through the photosphere per unit time. The decrease in the flux with increasing the differential rotation rate is due to the differential rotation lengthening polarity inversion lines in the east-west direction. The lengthened polarity inversion lines provide more locations where flux cancellation can occur, hence increasing the flux cancellation rates, and thus decreasing the overall flux on the surface of the star. Notably, the flux emergence rate has a greater effect on the surface flux than the differential rotation rate. This is evident when considering the diagonal in Figure \ref{flux_fig} where it can be seen that the surface flux has a net increase when both the flux emergence rates and differential rotation rates are increased by the same amount.

\begin{figure}
\centering\includegraphics[scale=0.5]{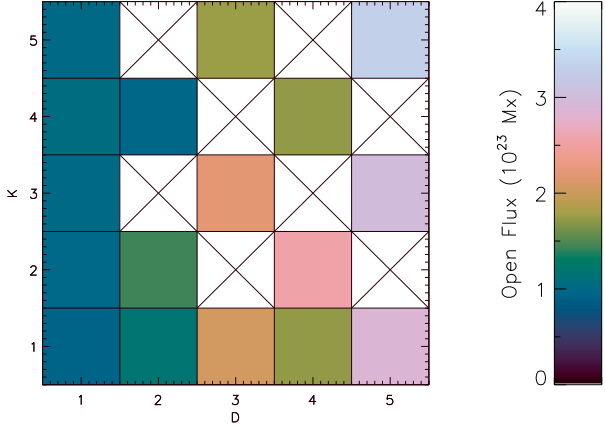}
\caption{The steady state open flux for each simulation. \label{opflux_fig}}
\end{figure}

\begin{figure}
\centering\includegraphics[scale=0.5]{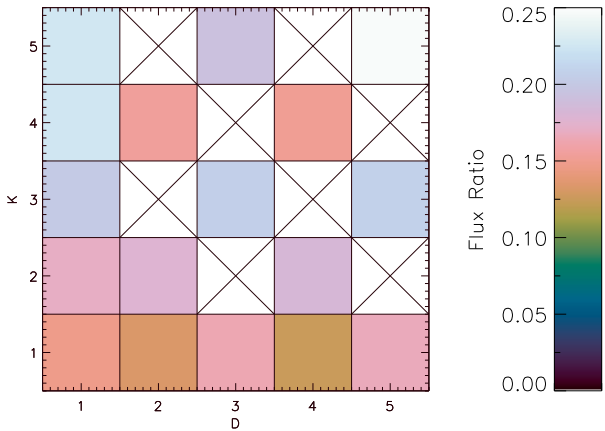}
\caption{The ratio of the open flux to the surface flux for each simulation \label{fratio_fig}}
\end{figure}

Figure \ref{opflux_fig} displays the open flux for each simulation. From this figure it can be seen that increasing the flux emergence rate increases the open flux, however increasing the differential rotation rate has little effect on the open flux. This is in contrast with the surface flux, where increasing the differential rotation rate decreases the flux. It is useful to consider the ratio of open flux to surface flux, which is displayed in Figure \ref{fratio_fig} for each simulation. This shows that the ratio of fluxes is reasonably insensitive to the flux emergence rate, yet {generally} increases with increasing differential rotation rate. This can be interpreted as the flux emergence rate having little effect on the overall distribution of flux in the corona (e.g. the total flux may increase, but the partition of open to closed flux does not). The differential rotation rate, however, does have a marked effect on the coronal structure. Increasing the differential rotation rate has the effect of opening up the corona -- increasing the ratio of open to closed field.
This opening up of the coronal field can be attributed to the differential rotation acting to convert the predominantly east-west aligned field of newly emerged bipoles to a more north-south alignment. This adds to the dipole moment of the star's global magnetic field, resulting in more open field.

\begin{figure*}
\includegraphics[scale=0.5]{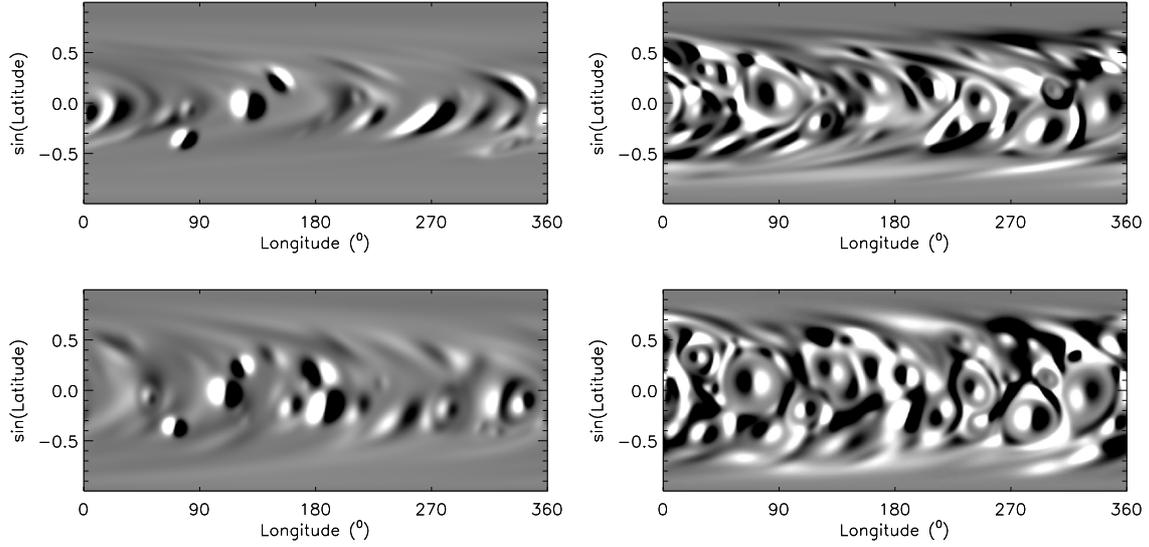}
\caption{{Synthetic} surface maps of the photospheric field on day 200 for the stars with $(D,K) = (1,5)$-- top left, $(5,5)$--top right, $(1,1)$--bottom left and $(5,1)$--bottom right. The images saturate at $\pm50$ G. \label{magnetogram_plots}}
\end{figure*}

\subsection{Distribution of Surface Field}

We now consider the distribution of magnetic flux on the stellar surface. Figure \ref{magnetogram_plots} displays synthetic magnetograms of the stellar surface for a range of different flux emergence and differential rotation rates. It is immediately apparent upon inspection that the stellar surface has become filled with active regions in the simulations with high flux emergence rates (right column of Figure \ref{magnetogram_plots}), compared to the simulations with the solar flux emergence rate (left column). The simulations with enhanced differential rotation (top row) appear to be slightly less filled with active regions than the simulations with solar differential rotation (bottom row). Also, the active regions in the simulations with enhanced differential rotation are elongated in the east-west direction compared to those in the simulations with solar differential rotation. As mentioned in the previous section, this is due to the differential rotation elongating features in the east-west direction. This lengthens the polarity inversion lines, allowing flux cancellation to occur more efficiently. As a result active regions have a shorter lifetime on high differential rotation stars, and thus at any given time these stars have fewer active regions than a star with an equivalent flux emergence rate, but a lower differential rotation rate.

\begin{figure}
\includegraphics[scale=0.5]{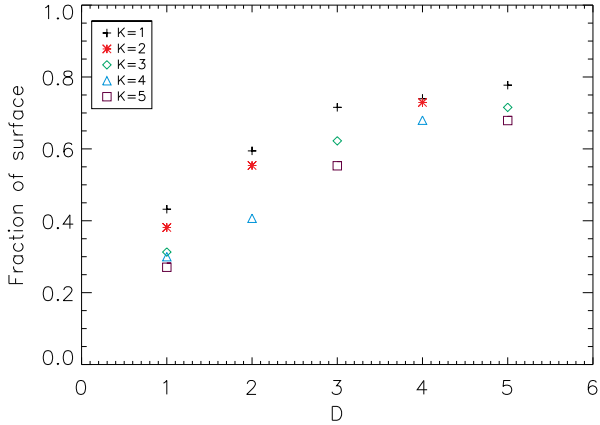}
\caption{Fraction of the photosphere at active latitudes covered with field stronger than 10 G for each simulation. For all differential rotation rates investigated the coverage increases with increasing flux emergence rates, and appears to be saturating at around 80\%. \label{arealats_fig}}
\end{figure}

Figure \ref{arealats_fig} displays the fraction of the active latitude bands ($\pm 40^\circ$ latitude) covered with field stronger than 10 G for each simulation. From the figure it can be seen that stars with a greater differential rotation rate have a lower fractional coverage of field. For every differential rotation rate investigated, it can also be seen that the coverage increases with increasing flux emergence rate, and seems to be saturating at around 80\% coverage. Whilst the coverage seems to saturate with increasing flux emergence rate, the surface flux (Figure \ref{flux_fig}) does not saturate. This suggests that whilst further increasing the flux emergence rate continues to add new flux into the photosphere, this new flux is emerged through existing flux rather than into `empty' regions of the photosphere. As such, the photosphere is saturated with active regions.

\begin{figure}
\includegraphics[scale=0.5]{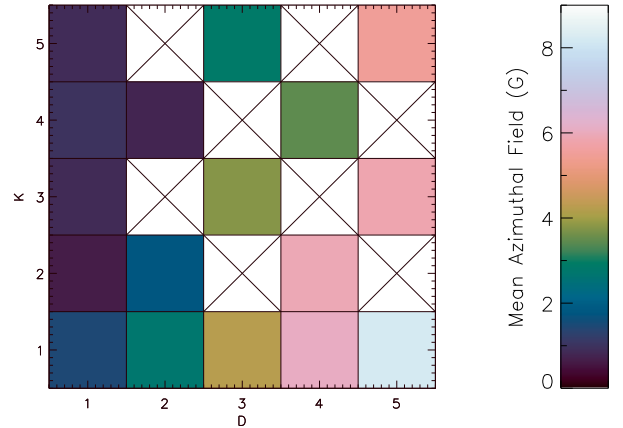}
\caption{Mean azimuthal field strength ($B_\phi$) for each simulation. \label{azimuthal_fig}.}
\end{figure}

With a number of ZDI studies showing strong rings of azimuthal field (e.g. \citet{Marsden2006,Donati1997,Donati22009}), we consider the mean azimuthal field present on the stars. To do this we calculate the mean photospheric azimuthal field ($B_\phi$) in each hemisphere. This is displayed in Figure \ref{azimuthal_fig}. It shows an increase of the azimuthal field with increasing flux emergence rate and possibly a slight decrease with increasing differential rotation rate. The increase with increasing flux emergence rate can be explained as follows. Each active region is emerged with an approximate east-west orientation. The horizontal field present in the active region is therefore predominantly aligned in the east-west, azimuthal direction. According to Hale's law \citep{Hale1919}, the leading polarity of each bipole in a given hemisphere tends to be the same sign over a given cycle, however the leading polarity is a different sign in each hemisphere. Therefore the majority of bipoles in a given hemisphere are aligned in the same direction, and so the horizontal fields of each bipole produce a net horizontal field in that hemisphere. Increasing the flux emergence rate increases the number of bipoles in each hemisphere, and thus increases the total horizontal field in the hemisphere. The slight decrease in mean azimuthal field strength with increasing differential rotation rate may be due to differential rotation taking an east-west aligned bipole and rotating it so the field is more north-south aligned, transforming some of the azimuthal field into poloidal field.

\subsection{Coronal Energetics}
\begin{figure}
\includegraphics[scale=0.5]{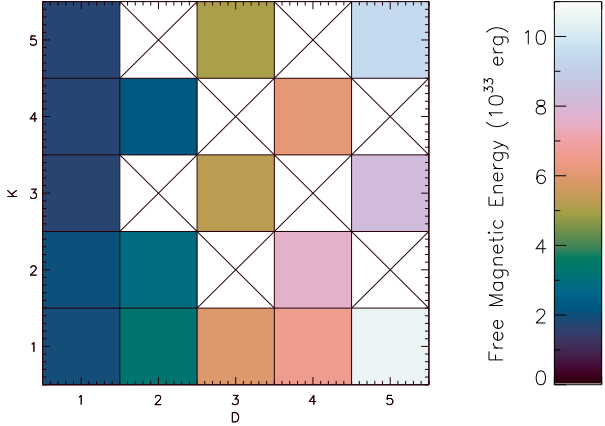}
\caption{Free magnetic energy for each simulation. The free magnetic energy increases with increasing flux emergence rate, however is insensitive to the differential rotation rate. \label{fenergy_fig}}
\end{figure}

\begin{figure}
\includegraphics[scale=0.5]{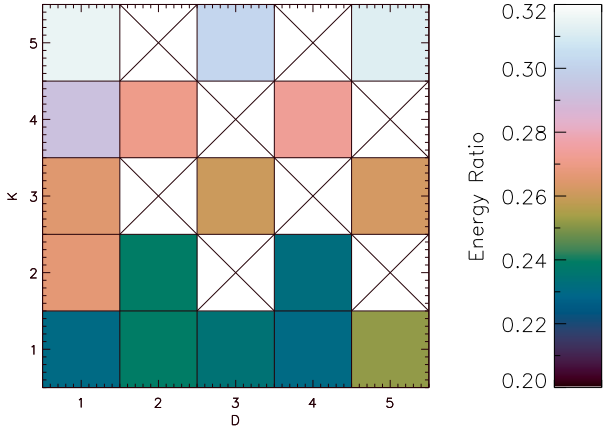}
\caption{Ratio of free magnetic energy to total magnetic energy for each simulation. The ratio is insensitive to the flux emergence rate, however increases with increasing differential rotation rate. \label{eratio_fig}}
\end{figure}

The photospheric evolution injects magnetic energy into the corona through a Poynting flux. The effects of flux emergence and differential rotation on the degree of energy injection is investigated in this section. Firstly, we consider the free magnetic energy, defined as
\begin{equation}
E_f = \frac{1}{8\pi} \int \left(B^2 - B_p^2\right) dV,
\end{equation}
where $B$ is the magnetic flux density of the non-potential field and $B_p$ is the magnetic flux density for a potential field with the same photospheric flux distribution as $B$. The free magnetic energy represents the maximum amount of magnetic energy that can be liberated, and thus represents the upper limit of energy available to drive flares and coronal mass ejections. Figure \ref{fenergy_fig} displays the steady state free magnetic energy for each simulation. It shows that the free magnetic energy increases with increasing flux emergence rate, yet seems to be reasonably insensitive to the differential rotation rate. It is also useful to consider the ratio of the free magnetic energy to the total magnetic energy. This measures how much the field deviates from a potential field. This quantity is displayed in Figure \ref{eratio_fig} and shows that the energy ratio increases with increasing differential rotation, but is insensitive to the flux emergence rate. This implies that whilst the differential rotation has little effect on the amount of free magnetic energy in the corona, it does result in a more non-potential corona. This increase in non-potentiality with increasing differential rotation rate is likely due to the increased differential rotation producing more shear in the corona, making it less potential.

\begin{figure}
\includegraphics[scale=0.5]{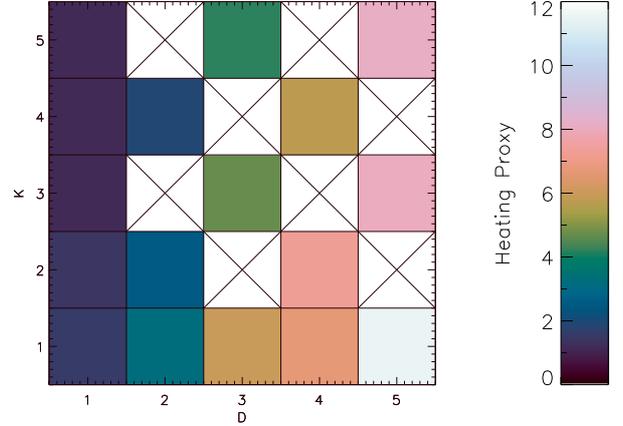}
\caption{Heating proxy for each simulation. The heating proxy increases with increasing flux emergence rate, and decreases with increasing differential rotation rate. \label{j2_fig}}
\end{figure}

One of the sources of heating in the solar corona is Ohmic heating, which is proportional to $j^2$. Although in our simulations we do not explicitly have any Ohmic diffusion, by calculating the volume integrated $j^2$ we may obtain a proxy for the heating present in the corona. Figure \ref{j2_fig} displays the volume integrated $j^2$ in the simulations. Similar to the flux, the heating proxy increases with increasing flux emergence rate, and decreases with increasing differential rotation rate. The flux emergence rate has a greater effect on this proxy than the differential rotation rate, however, as increasing both the flux emergence rates and differential rotation rates results in an increase in the heating proxy.

\subsection{Flux Ropes} \label{proj3_fr}
We now investigate the flux ropes in the stellar coronae. There are two methods that we use to detect flux ropes, developed in \citet{Yeates2009} and \citet{Gibb2014-2}. The first method, hereafter the `forces' method, considers the Lorentz force in the corona. At the axis of a flux rope, the magnetic pressure force is directed radially out away from the axis, whilst the magnetic tension force is directed radially inwards towards the axis. Flux ropes can thus be located by considering locations where these force criteria are met. The second method we use, hereafter the `PIL' method, considers the angle of the horizontal field component across a polarity inversion line (PIL) at the photosphere. If the angle between the normal to the PIL and the horizontal field component (shear angle) is greater than $90^\circ$ then this is indicative of a flux rope lying above the PIL. These two flux rope detection methods are described in detail in \citet{Gibb2014-2}.

\subsubsection{Non-erupting flux ropes}

First, we consider the number of non-erupting flux ropes present. In \citet{Gibb2014-2} it was found that a proxy for the onset of a flux rope eruption was the sudden disappearance of the length along a PIL where the shear angle is greater than $90^\circ$. Due to the magnetofrictional method an erupting flux rope rises slowly through the corona until it leaves the simulation though the upper boundary. The time between the onset of the eruption and the flux rope exiting the computational domain can be several days. During this time, although the erupting flux rope cannot be detected by the PIL method (as it is no longer in contact with the photosphere) it can still be detected by the forces method. Therefore, if we were to use the forces method to count the flux ropes in the corona, we would obtain the total number of erupting and non-erupting flux ropes. We thus use the PIL method to count the non-erupting flux ropes, as this method is unable to detect erupting flux ropes.

\begin{figure}
\centerline{\includegraphics[scale=0.49]{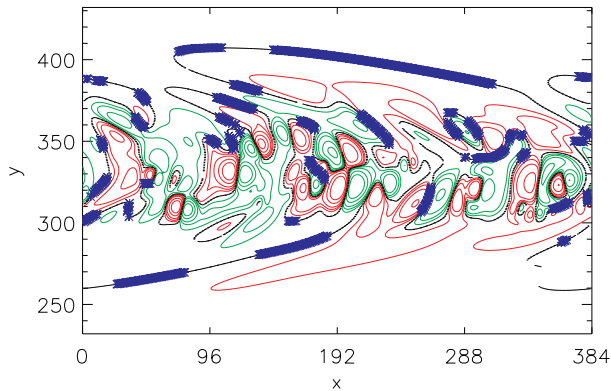}}
\caption{Flux rope locations (blue crosses) as determined by the PIL method for the simulation with $(D,K)=(1,1)$ on day 193. The green and red contours denote negative and positive magnetic flux respectively. The grey lines denote the polarity inversion lines. \label{proj3_emerg_pilfinder}}
\end{figure}

Figure \ref{proj3_emerg_pilfinder} displays the flux rope locations determined by the PIL method for the simulation with the solar flux emergence rate on day 193 of the simulations. As can be seen, there are many points belonging to flux ropes present. In order to count the number of flux ropes, we apply the following algorithm:

\begin{enumerate}
\item Flux rope points with fewer than 2 neighbours within a 5 grid cell radius are discarded. This is to remove small regions which could either be false positives or the residual areas of shear angle greater than $90^\circ$ which can occur just after a flux rope has erupted.

\item We then construct a 2-dimensional array with one element per grid cell (in the $x$ and $y$ directions). For grid cells with flux ropes present, the corresponding elements in the array are set to one. Elements corresponding to grid cells without flux ropes are set to zero.

\item We then smooth this array with a smoothing width of 5 grid cells. This acts to smear out and `join up' the discrete flux rope points to make one continuous region. Points in the array that are non-zero are set to 1.

\item The array is then contoured, and the number of flux ropes is taken to be the number of closed contoured regions.
\end{enumerate}

This method, though generally good at determining the number of flux ropes has some shortcomings. Firstly, if the region of shear angle greater than $90^\circ$ along the length of a flux rope has a gap of more than 5 grid cells, then the algorithm counts two flux ropes instead of one, thus overestimating the number of flux ropes. An example of this is visible in Figure \ref{proj3_emerg_pilfinder} where the two northern-most flux ropes in fact belong to the same flux rope (according to field line plots). Another shortcoming is that if two distinct flux ropes are in close proximity to each other, the smoothing process may merge the two flux ropes into one object, thereby causing the algorithm to underestimate the number of flux ropes. Although these shortcomings exist, upon comparing the number of flux ropes as determined by the algorithm with the number of flux ropes counted by eye for several randomly chosen times, the two numbers were in agreement to within an error of 10\%.

\begin{figure}
\centering\includegraphics[scale=0.5]{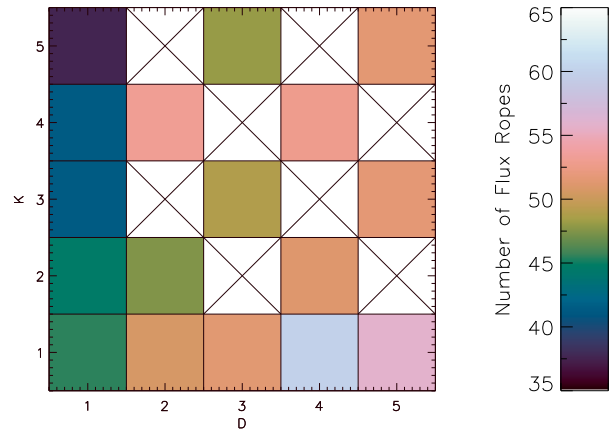}
\caption{The steady state number of flux ropes present on the stars in each simulation. \label{proj3_emerg_nfluxropes}}
\end{figure}

As with the flux, free magnetic energy, heating proxy etc, the number of flux ropes present reaches a steady state.
In Figure \ref{proj3_emerg_nfluxropes} the steady state number of flux ropes present on the stars at any one time is plotted for each simulation. The figure shows that the number of flux ropes present generally increases with increasing flux emergence rate and decreases with increasing differential rotation rate. The increase in the number of flux ropes with increasing flux emergence rate can be attributed to there being more active regions present, and so more polarity inversion lines where flux ropes may form. Considering the simulations with solar differential rotation, the number of flux ropes peaks at four times the solar flux emergence rate, then decreases. This could be due to the increasing flux emergence rate reducing the longevity of the flux ropes due to flux emergence events triggering flux rope eruptions. From Figure \ref{arealats_fig} it is clear that the photosphere is saturated with active regions' field, so new active regions have to emerge into pre-existing active regions. These emergences can disrupt the formation of flux ropes in the pre-existing active region, thereby also decreasing the number of flux ropes present.
The decrease in flux rope numbers with increasing differential rotation rates may be attributed to the lifetimes of flux ropes decreasing with increasing differential rotation rate, as found in \citet{Gibb2014-2}. In the following section we will estimate the flux rope lifetimes.
It should be noted that in each simulation, flux ropes only begin to be counted in each simulation a number of days after the simulation had started. The number of days taken for the first flux ropes to be counted was equal to the flux rope formation timescales found in \citet{Gibb2014-2}.

\subsubsection{Erupting flux ropes}

We now consider the rate of flux rope eruptions in each simulation. Since the number of flux ropes in the corona per day reaches a steady state, the number of flux ropes that form per day must be roughly equal to the number of flux ropes that erupt per day. Therefore, by determining the average eruption rate, we also determine the average formation rate. Considering that the mean number of flux ropes present per day on the star varies between 35 and 60 (depending upon the flux emergence and differential rotation rate), keeping count of the appearance and disappearance of flux ropes as determined by the PIL method is difficult. We therefore instead use the forces method to locate flux rope eruptions. To do this, we use a method based upon that developed by \citet{Yeates2009}, which involves looking for flux ropes that are moving upwards with $v_z \ge 0.5 \text{ km s}^{-1}$. As previously stated, flux ropes may take several days to completely leave the computational box once they have `erupted.' On any two consecutive days, the force method may thus identify the same erupting flux rope. We therefore are interested in the number of `new' erupting flux ropes on each day. As with the PIL method to count flux ropes described above, the method to count erupting flux ropes has some shortfalls. Firstly, two flux ropes that erupt close to each other (spatially and temporally) may be considered to be the same flux rope. Secondly, if one flux rope fragments into two distinct structures when erupting it may be counted as two flux ropes.

\begin{figure}
\centering\includegraphics[scale=0.5]{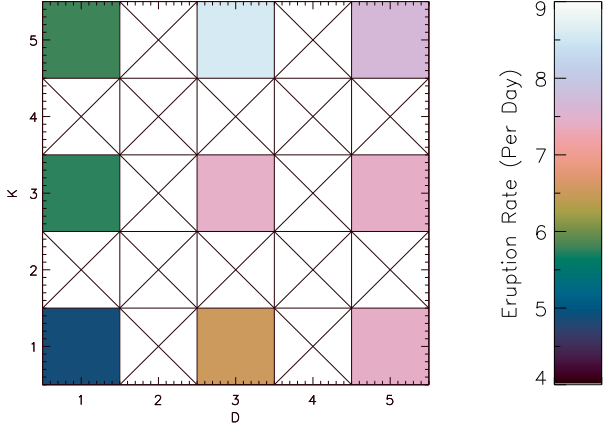}
\caption{The mean eruption rates for the simulations investigated. \label{eruptions_fig}}
\end{figure}

For a 60 day period (days 200--260) in a selection of the simulations we count the number of erupting flux ropes, and the eruption rate for each simulation is taken to be the mean number of eruptions per day.
Figure \ref{eruptions_fig} displays the determined rates. The eruption rates increase slightly with increasing differential rotation rate. This finding agrees with the speculations of \citet{Gibb2014-2}, who postulated that the flux rope eruption rates may increase with increasing differential rotation rate. This increase is small, however.
The eruption rates are also found to generally increase with increasing flux emergence rate.

\begin{figure}
\centering\includegraphics[scale=0.5]{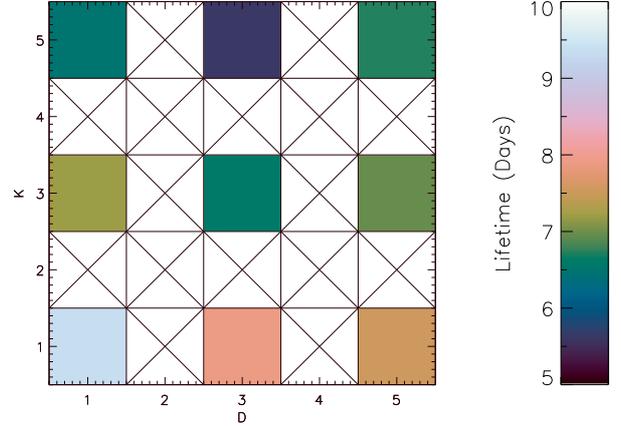}
\caption{The mean lifetimes of flux ropes in the simulations. \label{lifetimes_fig}}
\end{figure}

We may also crudely estimate the mean lifetime of flux ropes. We know that the mean formation rate is roughly equal to the mean eruption rate. So in any one day, on average $n_f$ flux ropes form, and $n_e=n_f$ erupt, keeping the total number of flux ropes in the corona, $N$, constant. If each flux rope has a mean lifetime of $l$ days, and every day $n_f$ flux ropes form, then the total number of flux ropes in the corona at any one time is approximately
\begin{equation}
N \approx n_f l
\end{equation}
and therefore the lifetime, $l$, can be approximated by $N/n_f$. It is important to compare these results with the findings of \citet{Gibb2014-2}, who found that the lifetimes of flux ropes on stars with 1, 3 and 5 times the solar differential rotation rate were 16, 6 and 4 days respectively. Figure \ref{lifetimes_fig} displays the lifetimes for the simulations where we calculated the eruption rates. The lifetimes determined vary between 5.6 and 9.34 days. This range is much smaller than the range found by \citet{Gibb2014-2}. The lifetimes vary with flux emergence rate, however a general trend cannot be seen. The lifetimes are found to decrease with increasing differential rotation rate, which is in agreement with \citet{Gibb2014-2}.
The range of 5.6--9.34 days found in the present study is lower than the range of 4--16 days found in \citet{Gibb2014-2}, implying that the addition of external coronal field weakens differential rotation's effect on the longevity of flux ropes.

\subsection{Emission Proxy Images of the Stellar Corona} \label{proj3_emerge_emission}
We now present emission proxy images of the stellar corona. This is achieved in a manner similar to that used in \citet{Gibb2014}. In order to construct these, we construct a simple atmospheric model where we assume that the emission at each point in the corona is proportional to the ohmic heating ($j^2$) and include a simple description of the plasma density. This density dependence is motivated by the fact that a portion of the corona with a higher density will emit more radiation as there are more particles present to emit photons. It is very important to note that our simulations do not include any plasma, so the density profile we use is unrelated to the coronal model and is only used to construct the emission proxy. By assuming an isothermal corona, we may obtain a crude description of the density as a function of height as
\begin{equation} \label{proj3_scaleheight}
\rho(h) \propto \exp\left(-\frac{h}{\Lambda}\right),
\end{equation}
where $h=r-R_*$ is the height above the photosphere and $\Lambda$ is the pressure scale height, given by
\begin{equation}
\Lambda= \frac{k_BT}{\bar{\mu} g},
\end{equation}
where $g$ is the surface gravity of the star (taken to be equal to the surface gravity of the Sun), choosing $T=2$ MK and $\bar{\mu} = 0.5m_p$ (where $m_p$ is the proton mass) we find $\Lambda = 120$ Mm.

\begin{figure*}
\centering\includegraphics[scale=2.5]{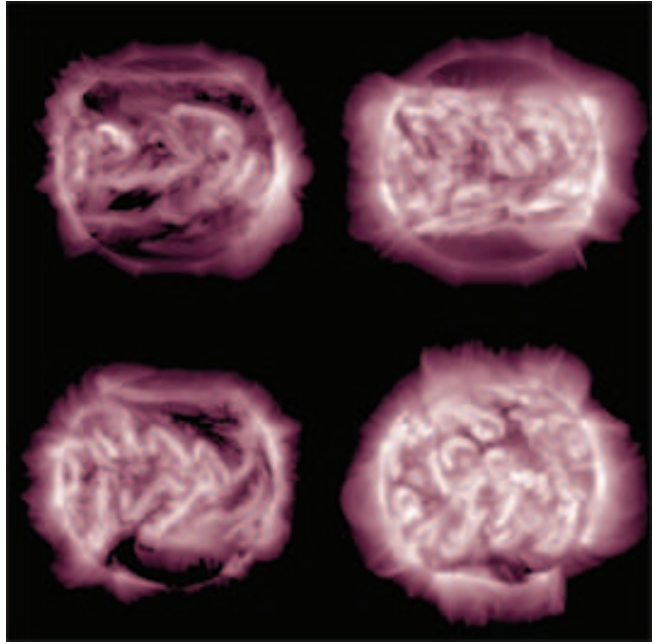}
\caption{{Synthetic} emission proxy images on day 200 of the simulations with $(D,K)=(1,5)$--top left, $(5,5)$--top right, $(1,1)$--bottom left and $(5,1)$--bottom right. The emission proxy images display the natural logarithm of the line-of-sight integrated emission proxy, and the colour scalings are identical in each image. \label{proj3_xray}}
\end{figure*}

\begin{figure*}
\centering\includegraphics[scale=2.0]{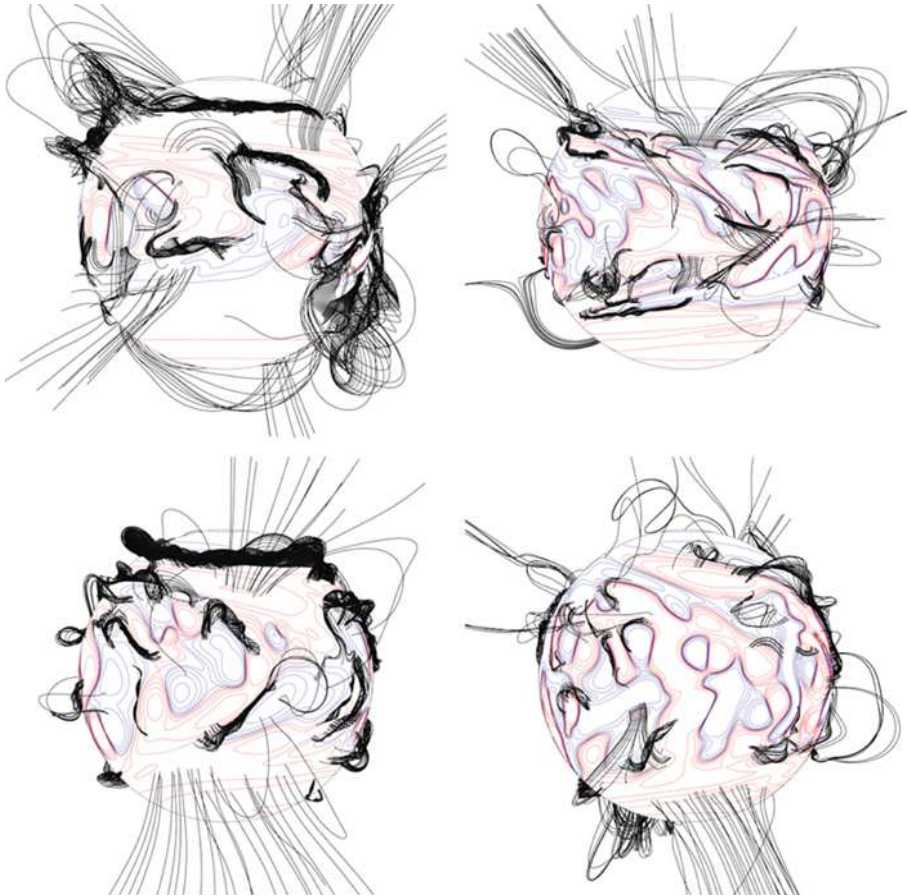}
\caption{Field line plots corresponding to the emission proxy images shown in Figure \ref{proj3_xray}. Red and blue contours denote positive and negative magnetic field respectively, and the black lines are field lines. \label{proj3_xray2}}
\end{figure*}

To construct our emission proxy images, we calculate $\rho^2(h) j^2$ within the coronal volume, and then integrate it along a line of sight to produce the image, using the assumption of optically thin radiation. This method does not represent the physical processes of radiative transfer, which must include the temperature of the plasma and its composition. As we do not have any information on the coronal plasma from the simulations, we cannot address the emission problem properly, and must instead use a proxy such as the one described above.

Figure \ref{proj3_xray} displays our emission proxy images on day 200 for stars with the solar flux emergence rate (left column), five-times the solar flux emergence rate (right column), five times solar differential rotation (top row) and solar differential rotation (bottom row). The images look remarkably similar to X-ray/EUV images of the hot solar corona taken by observatories such as SOHO/EIT, SDO/AIA and Hinode/XRT. In particular, active regions appear as bright loop structures, and unipolar regions (coronal holes) appear dark. The simulations with increased flux emergence rates (right column) appear brighter as they have many more active regions than the simulations with the solar flux emergence rate. On the simulations with high differential rotation (top row) the emission is mostly constrained to lower latitudes, with little emission at the poles. This is in contrast to the simulations with solar differential rotation (bottom row) which have sources of emission at much higher latitudes. This is due to the enhanced differential rotation increasing the dipole moment of the global magnetic field, resulting in more open polar regions. Figure \ref{proj3_xray2} displays field line plots for the same simulations and times as Figure \ref{proj3_xray}.

\section{Discussion and Conclusions} \label{discussion}
In this paper we have investigated the coronal response to varying the flux emergence and differential rotation rates. In order to do this we used a surface flux transport model to determine the evolution of the photospheric field, and used this to drive a non-potential coronal evolution via the magnetofrictional method. We use a flux emergence model that emerges simplified bipolar active regions onto the photosphere and into the corona. The bipole emergence model is based upon observations of the emergence of bipolar active regions on the Sun, however through varying a single parameter we could have the rate of emergences per day increased to simulate higher flux emergence rates. We carried out simulations with various values of the flux emergence and differential rates so that a parameter study could be carried out. The flux emergence rates were varied between one and five times the solar flux emergence rate of 0.62 bipoles per day, and similarly the differential rotation rates were varied between one and five times the solar differential rotation rate of $3.65^\circ$ per day.

Firstly we considered the effects of the flux emergence and differential rotation rates on the surface flux and open flux. It was found that the surface flux (and mean surface field strength, $\avg{B}$) increased with increasing flux emergence rate. Increasing the differential rotation rate had the effect of slightly decreasing the amount of surface flux. This was attributed to the increased differential rotation lengthening polarity inversion lines, allowing flux cancellation to occur more rapidly. Overall the flux emergence rate had a greater effect on the flux than the differential rotation rate. The open flux was found to increase with increasing flux emergence rate, however no trend was found between differential rotation rate and the open flux. In contrast, the ratio of open to surface flux was found to be independent of the flux emergence rate, but increased with increasing differential rotation rate. This means that increasing differential rotation alters the structure of the corona (opening it up) whilst though the flux emergence adds more flux to the corona, it does not alter the global coronal structure. This opening up of the corona with increasing differential rotation is due to the differential rotation converting east-west field into north-south field, adding to the star's overall dipole moment.

We next considered the response of the surface fields to flux emergence and differential rotation, namely the proportion of the photosphere covered with flux, and the mean azimuthal surface field strength. For all differential rotation rates investigated, it was found that the proportion of the active latitude bands ($\pm 40^\circ$ latitude) covered in field stronger than $10$ G increased with increasing flux emergence rate, and seemed to be saturating at around 80\% for high flux emergence rates. This implies that the surface becomes saturated with active regions for flux emergence rates of four to five times the solar rate. As the differential rotation rate was increased the fractional coverage slightly decreased, due to the lengthening of polarity inversion lines and more efficient flux cancellation as mentioned above. We also investigated the mean azimuthal field strength (per hemisphere) as a function of differential rotation rate and flux emergence rate. This is of interest observationally as a number of ZDI studies of stars have found rings of azimuthal field \citep{Donati1997,Marsden2006,Donati22009}. We find that the mean azimuthal field increases with increasing flux emergence rate and decreases with increasing differential rotation rate. As with the surface flux, the flux emergence rate has a greater effect on the azimuthal field compared to the differential rotation rate. The increase of the azimuthal field with increasing flux emergence rate is due to the increased number of east-west aligned bipoles contributing to the azimuthal field. The decrease in azimuthal field with increasing differential rotation rate is due to the differential rotation converting some of the azimuthal field into poloidal field by effectively rotating the bipoles to a more north-south alignment. Azimuthal rings of field seen in ZDI studies may be the result of a ring of densely packed active regions, whose combined azimuthal field appears as a ring in ZDI maps.

The effects of the flux emergence rate and differential rotation rate on the coronal energetics were also studied. We considered the free magnetic energy, the ratio of free magnetic energy to total magnetic energy, and the global currents. The amount of free magnetic energy was found to increase with increasing flux emergence rate, but is insensitive to differential rotation rate. The total energy available to power phenomena such as CMEs and flares is therefore affected by the flux emergence rate. The ratio of the free magnetic energy to the total magnetic energy, which is a measure of how much the field deviates from a potential field, shows the opposite behaviour. It was found to be insensitive to the flux emergence rate, but increase with differential rotation rate. Although differential rotation does not have an effect on the total amount of free magnetic energy in the corona, it does have an effect on the non-potentiality of the corona. This is likely due to the differential rotation adding more shear into the corona. Given that the degree of non-potentiality increases with increasing differential rotation rate, potential field extrapolations are likely to be most accurate for stars with low differential rotation.

We then considered the volume-integrated square of the current. This may be considered a proxy for the amount of heating in the corona, as Ohmic heating is proportional to $j^2$. This was found to increase with increasing flux emergence rate, and decrease with increasing differential rotation rate. As is the case with many of the quantities described above, this is more sensitive to the flux emergence rate than the differential rotation rate. Assuming a link between heating proxy and X-ray emission, the X-ray luminosities of stars are more affected by their flux emergence rates than their differential rotation rates, and stars with higher flux emergence rates may have higher X-ray luminosities.

In \citet{Gibb2014-2} the timescales of formation and eruption of flux ropes were determined as a function of differential rotation. It was postulated that stars with higher differential rotation rates may have more frequent flux rope eruptions, and thus may produce more CMEs. In order to address this hypothesis, we investigated the flux ropes formed in our simulations. Firstly, we determined the mean number of stable (non-erupting) flux ropes in the corona. This was found to increase with increasing flux emergence rate, and decrease with increasing differential rotation rate, but with a higher sensitivity to flux emergence rate than differential rotation rate. We then counted the mean number of flux ropes erupting per day for a number of the simulations. This eruption rate was found to increase with increasing flux emergence rate and with increasing differential rotation rate. We postulate that planets orbiting stars with differential rotation rates and/or flux emergence rates greater than those of the Sun may experience more frequent coronal mass ejections. Finally, we estimated the mean lifetimes of flux ropes. The lifetimes were found to decrease with both increasing flux emergence rate and increasing differential rotation rate. This decrease, however, was not as marked as was found in \citet{Gibb2014-2}.

A simple method to produce X-ray emission proxy images of the stellar corona was developed. This method, based upon the one employed by \citet{Gibb2014}, sets the emission per unit volume to be proportional to the square of the current density, with a weighting according to height above the photosphere -- motivated by the decreasing coronal density with height. The images produced look remarkably similar to X-ray/EUV images of the solar corona, displaying coronal holes and bright active regions. At present these images are only used as a way to visualise the coronae of the stars and are not used for any qualitative analysis, however in future it would be interesting to compare emission proxy images produced from global solar non-potential coronal field simulations (for example the fields of \citet{Yeates2014}) with actual X-ray/EUV observations of the Sun to determine their effectiveness.

In this study we used a flux emergence model based upon the observed solar flux emergences over a time period of a year. The flux emergence profiles on other stars may be different to the solar flux emergence profile. For example, the latitudinal distribution of emergences may be different, as may the tilt angles, or the mean flux per bipole. Although using a different flux emergence profile may provide different coronal properties, we feel that the trends with increasing flux emergence/differential rotation rates found in this paper will be roughly the same.

In addition to flux emergence and differential rotation, the coronal evolution is also influenced by surface diffusion and meridional flow. The coronal response to these effects have not been studied in this paper, however the effect of the surface diffusion on coronal timescales was investigated in \citet{Gibb2014-2}. Such a parameter study involving meridional flow and surface diffusion will be carried out in the future. One other effect which could be investigated is the height at which the coronal field is prescribed to be radial (in this work it is 2.5$R_*$). This may have an effect on the value of the open flux in the simulations, however we expect the trends displayed in Figures \ref{opflux_fig} and \ref{fratio_fig} would remain roughly the same.

In summary, we find that the flux emergence rate has a greater effect on a star's coronal properties than the differential rotation rate. The total surface flux, the total open flux, the mean surface field (both radial and azimuthal) the free magnetic energy and the heating all increased with increasing flux emergence rate. Each of these properties only showed a slight decrease with increasing differential rotation rate. Due to this it is extremely important that future observational studies aim to determine the rates and properties of (or suitable proxies for) flux emergence on other stars. Flux rope ejection rates increased with increasing flux emergence and differential rotation rates, which implies that planets orbiting such stars may be exposed to more frequent CMEs. This is important for the longevities of planetary atmospheres. Whilst flux emergence has a stronger effect on the coronal properties than the differential rotation, it was found that the differential rotation altered the coronal structure. As the differential rotation rate increases the corona becomes more open and non-potential.

\section*{Acknowledgements}
G.P.S.G. would like to thank the STFC for financial support. D.H.M. would like to thank the STFC and the Leverhulme Trust for financial support. Simulations were carried out on a STFC/SRIF funded UKMHD cluster at St Andrews.

\bibliographystyle{mnras}
\bibliography{refs}

\bsp	
\label{lastpage}
\end{document}